\documentclass[twocolumn,showpacs,preprintnumbers,amsmath,amssymb]{revtex4}

\usepackage{graphicx}
\usepackage{dcolumn}
\usepackage{bm}

\begin{document}


\title{Polarization and angular distribution of the radiation emitted in laser-assisted recombination}

\author{S. Bivona}
\author{G. Bonanno}%
\author{R. Burlon}%
\author{C. Leone}%
 \affiliation{%
Dipartimento di Fisica e Tecnologie Relative, Universit\'a degli Studi di Palermo - Italy and C.N.R. - C.N.I.S.M.\\
Viale delle Scienze - Ed. 18 - I-90128 Palermo Italy
}%

\email{bivona@unipa.it}

\date{\today}

\begin{abstract}
The effect of an intense external linear polarized radiation field on the angular distributions
and polarization states of the photons emitted during the radiative recombination is investigated.
It is predicted, on symmetry grounds, and corroborated by numerical calculations of approximate
recombination rates, that emission of elliptically polarized photons occurs when the momentum
of the electron beam is not aligned to the direction of the oscillating field. Moreover, strong modifications
to the angular distributions of the emitted photons are induced by the external radiation field.
\end{abstract}

\maketitle

The capture of a free electron by a positive ion is one of the
fundamental processes occurring in the electromagnetic interaction
of charged particles. The recombination of an electron with a bare
ion is made possible by the spontaneous emission of a photon, in
order to fulfill both energy and momentum conservation.

In view of its application to the analysis of astrophysical and
laboratory plasmas as one of the principal means to obtain
information about their physical conditions through the observation
of the emitted radiation, this process has been extensively studied
both theoretically and experimentally. In the last years much work
has been devoted to radiative recombination (RR) in the presence of
laser field~\cite{BetheSalpeter,Hahn}.

The availability of intense laser sources has stimulated theoretical
investigation of laser assisted radiative recombination (LARR) aimed
at exploring the spectrum of the radiation emitted during the
recombination event. In the presence of intense monochromatic
electromagnetic fields, the electron can exchange a large number of
photons of the assisting radiation field and may recombine in a
bound state~\cite{Kuchiev,Jaron,Leone} emitting x-rays whose frequency depends on
the number of the exchanged laser photons, according to the energy
conservation law:

\begin{equation}\label{Econservation}
\hbar\omega_X(n)=\frac{p^2}{2m}+I_f+\Delta+n\hbar\omega_L
\end{equation}

In Eq.~\ref{Econservation}, $\omega_X(n)$ is the frequency of the
emitted photon, $p^2/2m$ is the incoming electron energy, $I_f$ is
the ionization energy of the bound state in which the electron is
captured, $\Delta=e^2 E_0^2/4 m \omega_L^2$ is the ponderomotive
shift, $n$ denotes the number of exchanged photons, $\omega_L$ and
$E_0$ the frequency and the amplitude of the oscillating laser
field, $e$ and $m$ the electron charge and the mass respectively.

From Eq.~\ref{Econservation}), it may be easily seen that the
spectrum of the emitted x-ray results
in a series of frequencies evenly separated by the assisted field
frequency.

The main features of the spectra of the emitted radiation have been
found by several authors by treating the process in the framework of a
Keldysh-type approach, where the interaction of the incoming
electron with the ion in the initial continuum state is neglected~\cite{Kuchiev}.
Slight modifications of this scheme have been
considered in order to take into account, though approximately, the
electron-ion interaction. In all the proposed
treatments, the calculated emission spectra show a large plateau due
to absorption of a large number of photons, followed by an abrupt
cutoff~\cite{Jaron,Leone,Cerkic}.

As it is well known, in the field-free process, when the electron
recombines with a bare ion in the ground state, the matrix element
of the electron dipole moment is directed along the direction of
the incoming electron momentum, and the emitted photon turns out to
be linearly polarized. In the dipole approximation, the photon
polarization vector lies in the plane formed by the direction of the
electron initial asymptotic momentum ${\bf q}$ and the one of the
wavevector of the emitted photon ${\bf k_\gamma}$. Due to the
cylindrical symmetry about the axis passing through the nucleus,
directed along ${\bf q}$, the intensity of the emitted radiation
depends only on the angle between ${\bf q}$ and ${\bf k_\gamma}$.
Moreover, the linear polarization degree of the emitted radiation is
equal to one.

For the subsequent analysis concerning the photon polarization state
emitted in the presence of the radiation field it may result  more
suitable to look at the above result in terms of the symmetries of
the physical system. In fact, upon reflection in the plane
containing ${\bf q}$ and ${\bf k_\gamma}$ the system does not
change, while the photon helicity reverse its sign. Therefore the
emission probability of photons characterized by opposite helicities
and equal ${\bf k_\gamma}$ is the same, and the unitary value of the
degree of linear polarization of the photon emitted in the radiative
recombination follows from the property that a linearly polarized
state may be considered as a superposition of two circularly
polarized states with opposite helicities and equal amplitudes.

It is the aim of this paper to study the modifications of both the
polarization states and the angular distribution of the emitted
photons when the recombination occurs in the presence of a linearly
polarized laser. Due to the circumstance that the highest coupling
between the recombining electron and the assisting radiation takes
place when the oscillating electric field is directed along the
incoming electron momentum ${\bf q}$, all the previous study, to the
best of our knowledge, has been carried out mainly in this geometry.
As the cylindrical symmetry, in this configuration, is still
maintained, for each emission channel characterized by the number of
photons that the assisting field exchanges with the recombining
electron, the angular distribution of the emitted radiation depends
on the intensity of the external field, while the photon
polarization properties remain unaltered. By changing the angle
between the direction of the laser field polarization and ${\bf q}$,
the above symmetry breaks down, affecting both the polarization
states and the angular distribution of the emitted photons.

In order to make quantitative estimations of the external field
effects we have to calculate the recombination transition amplitude
of the electron with a bare ion of charge $Ze$ in presence of an
intense, spatially and temporally homogeneous, radiation field that
will be taken in dipole approximation and treated classically.
During the recombination emission of x-ray photon occurs with
simultaneous exchange of laser photons. The x-ray radiation,
characterized by the wavevector ${\bf k_\gamma}$ and the polarization
vector $\hat{\bm \epsilon}_{{\bf k_\gamma}, \lambda}$, will be
treated in second quantization  and in dipole approximation.
Accordingly, the electric field operator associated to a single mode
x-ray radiation is taken, in Gaussian units, as

\begin{equation}\label{field}
{\bf \hat{E}_X}(t)=\sum_{\lambda=1,2} i \sqrt{\frac{2\pi \hbar
\omega_X}{V}}{ \hat{\bm \epsilon}}_{{\bf k_\gamma}, \lambda}
(\hat{a}_{{\bf k_\gamma}, \lambda} e^{-i \omega_X t}-\hat{a}^+_{{\bf
k_\gamma}, \lambda} e^{i \omega_X t})
\end{equation}

with  $V$ the quantization volume of the radiation, $\hat{a}_{{\bf
k_\gamma}, \lambda}$ and $\hat{a}^+_{{\bf k_\gamma}, \lambda}$ the
annihilation and creation operator, respectively, for a photon in the
state characterized by ${\bf k_\gamma}$ and $\lambda$. The nonrelativistic
hamiltonian of the atomic system interacting with the radiation
field reads

\begin{equation}\label{hamiltonian}
\hat{H} = \hat{H}_{at} - e {{\bf \hat{E}}_L \cdot {\bf r}} - e {{\bf
\hat{E}}_X \cdot {\bf r}}
\end{equation}

where $\hat{H}_{at}$ is the field-free atomic hamiltonian
$\hat{H}_{at}= \hat{p}^2/(2m)+ Ze^2/r$,
${\bf r}$ is the electron coordinate and ${\bf E}_L(t)$ the external
oscillating electric field assumed linearly polarized and directed
along the unitary vector ${\hat{\bm \epsilon}_L}$, $E_L={ \hat{\bm \epsilon}_L}
E_0 cos (\omega_L t)$.

The transition amplitude of emitting one x-ray photon characterized
by ${\hat{\bm \epsilon}}_{{\bf k_\gamma}, \lambda}$ during the
recombination of an electron with asymptotic average momentum ${\bf
q}$ from a continuum state $\psi_q^+({\bf r},t)$ into a bound state
$\psi_0({\bf r},t)$, treating the electron-x-ray photon interaction
at the first order of the time-dependent perturbation theory, is
given by (hereafter atomic units will be used)

\begin{equation}\label{matrix}
T_{i,f}=- \sqrt{\frac{2 \pi \omega_X}{V}}\int_{-\infty}^\infty dt
\langle \psi_0({\bf r},t) | {\bm \epsilon}_{{\bf k_\gamma}, \lambda}
\cdot {\bf r} e^{i\omega_X t}|\psi_q^+({\bf r}, t) \rangle
\end{equation}

By assuming {the oscillating laser field amplitude $E_0$ to be much
less than the interatomic electric field experienced by the electron
in the ground state of the hydrogenic ions,  $\psi_0({\bf r},t)$ may
be approximated by the field-free ground state of the hydrogenic ion
with charge $Z-1$ and energy $Z^2I_0$, where $I_0$ denotes the
hydrogen ground state energy.

Below, the continuum electron state $\psi_{\bf q}^+$ will be
approximated by the Coulomb-Volkov ansatz~\cite{Jain,Leone2,Zhang}
$\psi_{q}^+({\bf r},t)=\chi_{\bf q}({\bf r},t) u_q^+({\bf r})$
with
\begin{equation}\label{ChiVolkov}
\chi_{\bf q}=exp \{i[{\bf k_L}(t) \cdot {\bf r}-\frac{1}{2}\int^t dt'
[{\bf q}+{\bf k_L}(t)]^2]\},
\end{equation}
${\bf k_L}(t)$ the quiver momentum imparted to the electrons by
the intense field, ${\bf k_L}(t)={\hat{\bf \epsilon}_L} (E_0/\omega_L) \sin
\omega_L t$, and

\begin{equation}\label{Coulomb}
u_{\bf q}^+ = exp(1/2 \pi \nu) \frac{\Gamma(1-\nu)}{(2 \pi)
^{\frac{3}{2}}} {_1F_1}[i \nu, 1, i({\bf q \cdot r}-qr)] \exp(i{\bf q \cdot r})
\end{equation}

the field-free outgoing Coulomb wave ($\nu=Z/q$).

Proceeding in the usual way, by using the Fermi golden rule, the
differential recombination probability per unit time is obtained as a sum of
probabilities of single events in which the emission of an x-ray
photon occurs simultaneously to exchange of $n$ laser photons

\begin{equation}\label{somma}
\frac{dP}{d\Omega}=\sum_{n;\lambda=1,2} P_{n}(\hat{\bm \epsilon}_{{\bf k_\gamma,\lambda}})
\end{equation}.

In Eq.~(\ref{somma}) $d\Omega$ is the element of solid angle surrounding
the vector ${\bf k_\gamma}$ and $P_n(\hat{\bm \epsilon}_{{\bf k_\gamma,\lambda}})$
is the single-channel
differential rate of recombination with exchange of $n$ laser
photons and emission of an x-ray photon with energy $\hbar
\omega_X(n)$ and polarization vector $\hat{\bm \epsilon}_{{\bf
k_\gamma,\lambda}}$ with $\omega_X(n)$ given by
Eq.~(\ref{Econservation})

\begin{equation}\label{Prob}
P_n(\hat{\bm \epsilon}_{{\bf k_\gamma,\lambda}})=\int_0^\infty d\omega
\frac{\omega^3}{2 \pi c^3}|T_n(\hat{\bm \epsilon}_{{\bf k_\gamma,\lambda}})|^2
\delta[\omega-\omega_X(n)]
\end{equation}

with $T_n(\hat{\bm \epsilon}_{{\bf k_\gamma,\lambda}})$ the corresponding
transition amplitude given by

\begin{widetext}
\begin{equation}\label{transition}
T_n(\hat{\bm \epsilon}_{{\bf k_\gamma},\lambda})=\frac{1}{2 \pi}
\int_{-\pi}^\pi d\alpha~exp\{i[n\alpha+{\bf q \cdot
\hat{\epsilon}_L}\frac{k_L}{\omega_L} \cos \alpha+\frac{k_l^2}{8
\omega_L^2} \sin 2 \alpha]\}\times \langle u_0({\bf
r})|\hat{\epsilon}_{{\bf k_\gamma,\lambda}} \cdot {\bf r} | e^{i{\bf k_L
\cdot r} \sin \alpha} u_{\bf q}^+({\bf r}) \rangle
\end{equation}
\end{widetext}

\begin{figure}[htbp]
\centering{\resizebox{5cm}{!}{\includegraphics{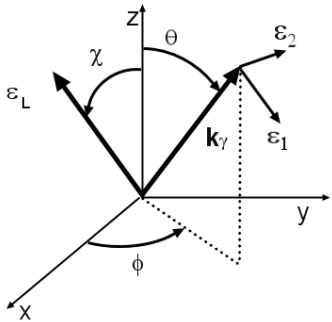}}}
\caption{\label{frame} The coordinate system used to calculate the differential recombination
rate. The asymptotic electron momentum is directed along $z$; the emitted photon wavevector
${\bf k_\gamma}$ points into the direction $(\theta, \phi)$; the oscillating electric field,
lying in the plane $xz$ makes an angle of $\chi$ with the $z$ axes.}
\end{figure}

As shown in Fig.~\ref{frame}, the emission direction of the x-ray
photon is specified by the angles $\theta$ and $\phi$, while the angle
between  $\hat{\bm \epsilon}_L$ and ${\bf \hat{z}}$, the unit vector
indicating the incoming electron momentum direction (${\bf q} = {\bf
\hat{z}} q$), is denoted by $\chi$. Without loosing generality, let us
denote by $\hat{\bm \epsilon_1}(0)\equiv \hat{\bm \epsilon}_{{\bf k_\gamma},1}$ the
linear polarization vector lying in the plane containing  ${\bf
k_\gamma}$ and ${\hat{\bf z}}$, and by $\hat{\bm \epsilon}_2(0)\equiv \hat{\bm \epsilon}_{{\bf
k_\gamma},2}$ the one lying in the $xy$ plane, so that
$(\hat{\bm \epsilon}_1(0),\hat{\bm \epsilon}_2(0),{\bf k_\gamma}/|{\bf k_\gamma}|)$ form
a right-handed set of mutually orthogonal unit vectors.

Upon replacement, in Eq.~(\ref{transition}), of $\hat{\bm \epsilon}_{{\bf k_\gamma},\lambda}$
with $\hat{\bm \epsilon}_{\pm 1}$ defined by
$\hat{\bm \epsilon}_{\pm 1}= \mp [{\bf \hat{\bm \epsilon}_1}(0) \pm
i {\bf \hat{\bm \epsilon}_2}(0)]/\sqrt{2}$,
the emission rate of an x-ray photon with helicity equal, respectively, to $+1$ or $-1$
is obtained. Below we characterize the polarization  state of the
emitted photon through the so called Stokes parameter $S_0$, $S_1$,
$S_2$, $S_3$~\cite{Born}. These are defined in terms of four
independent set of measurements giving information about the total
intensity of the x-ray radiation and the intensity transmitted by a
polarizer oriented at particular angles that will be specified
below.

The parameters $S_0\equiv I_r \propto
(P_n[\hat{\bm \epsilon}_1(0)]+P_n[\hat{\bm \epsilon}_2(0)])=P_n$ is given by the total
intensity of the radiation of a given frequency emitted in a
particular direction. Usually, the following normalized parameters
$\eta_i=S_i/S_0$ ($i=1,2,3$) are used.
$\eta_1=\{P_n[\hat{\bm \epsilon}_1(0)]-P_n[\hat{\bm \epsilon}_2(0)]\}/P_n$ gives the
degree of linear polarization with respect to the axes along
${\hat{\bm \epsilon}_1}(0)$ and ${\bf \hat{\bm \epsilon}_2}(0)$. It is obtained by
measuring the radiation intensity $I[{\dot{\hat{\bm \epsilon}}}_i(0)]
\propto P_n[{\hat{\bm \epsilon}}_i(0)]$ transmitted by a polarization
filter oriented along the direction of ${\hat{\bm \epsilon}}_1(0)$ and
${\hat{\bm \epsilon}}_2(0)$;
$\eta_2=\{P_n[\hat{\bm \epsilon}_1(\pi/4)]-P_n[\hat{\bm \epsilon}_2(\pi/4)]\}/P_n$ is
the degree of linear polarization with respect to two orthogonal
axes rotated by $\pi/4$ with respect to the above ones. The degree
of circular polarization  is given by
$\eta_3=(I_+-I_-)/(I_++I_-)=\{P_n[\hat{\bm \epsilon}_+]-P_n[\hat{\bm \epsilon}_-]\}/P_n$
with $I_+ \propto P_n({\hat{\bm \epsilon}_+})$ and $I_- \propto P_n({\bf
\hat{\bm \epsilon}_-})$ the intensities transmitted by polarization filters
that transmit photons with positive or negative helicity,
respectively. We remark that for the case under study, the radiation
is completely polarized and, hence, $\eta_1^2+\eta_2^2+\eta_3^2=1$.

By using Eq.~(\ref{transition}), calculation of energy spectra, angular distributions and
degrees of polarization of the emitted photons have been carried out at
moderate laser intensities, $I_L=3 \cdot 10^{13}~W/cm^2$, for such incoming
electron energies and nuclear charges that the requirements of radiation dipole
approximation and non relativistic treatments are satisfied.

In Fig.~\ref{segnali} the effect of the disalignment between
${\hat{\bm \epsilon}_L}$ and ${\bf q}$ on the angular distribution of the emitted x-ray
for selected channels characterized by the number of the exchanged assisting-field photons
is illustrated by plotting, as a function of $\theta$, the differential recombination rate
$P_n[\hat{\bm \epsilon}_1(0)]$ evaluated for three different values of the angle $\chi$, keeping fixed at
$45^\circ$ the value of the angle $\phi$.
The incoming electron energy has been taken equal to $60~eV$.
 The unusual behavior of the differential emission rate as a function
of the angle $\theta$, when $\chi \neq 0$, is quite remarkable.

\begin{figure}[htbp]
\centering{\resizebox{7.5cm}{!}{\includegraphics{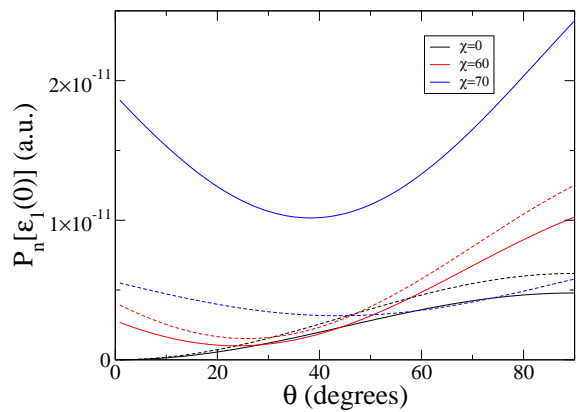}}}
\caption{\label{segnali} Differential recombination rate $P_n[\hat{\bm \epsilon}_1(0)]$,
with exchange of $n$ laser photons,
evaluated at three different values of $\chi$, as a function of the angle $\theta$ between ${\bf k_\gamma}$ and ${\bf q}$.
${\bf k_\gamma}$ points in the direction ($\theta$,$\phi$) with $\chi$ kept at $45^\circ$. Full line $n=25$; dashed line $n=28$.
}
\end{figure}

In the field-free recombination, or when the oscillating assisting electric field is
directed along the direction of the incoming electron momentum, the emission of
photon along {\bf q} is strictly forbidden. In fact, the electron states involved in
the transition are characterized by the same magnetic quantum number $m=0$,
and photon emission along ${\bf q}$ would violate the angular momentum conservation.
Instead, if ${\bf q}$ and ${\bf \bm \epsilon_L}$ are not aligned, $m$ ceases to be a good
quantum number, and the above restriction on the emission direction no longer applies.

Moreover, as already mentioned, the lack of cylindrical symmetry affects the polarization
states of the emitted photons. This is shown in Fig.~\ref{stokes} where together with
the differential emission rate of x-ray photons into different channels characterized by $n$,
for a fixed emission direction, are displayed the degrees of polarization $\eta_1$,
$\eta_2$ and $\eta_3$ as a function of $\chi$. We note that,
as $\eta_1^2+\eta_2^2+\eta_3^2=1$, the highest degree of circular polarization
occurs in proximity of values of $\chi$ at which both the linear polarization degrees
$\eta_1$ and $\eta_2$ reach their minimum absolute value.
The circular polarization degree may be written, by using Eqs.~(\ref{Prob},\ref{transition}), as

\begin{equation}\label{eta3}
\eta_3=\frac{4 \cdot Im \{T_n[{\bf \hat{\bm \epsilon}_1}(0)] \cdot T_n^*[{\bf \hat{\bm \epsilon}_2}(0)]\}}
{|T_n[{\bf \hat{\bm \epsilon}_1}(0)]|^2 + |T_n^*[{\bf \hat{\bm \epsilon}_2}(0)]|^2}= \sin 2 \psi
\end{equation}

with $\psi$ the ellipticity angle of the emitted radiation.

\begin{figure}[htbp]
\centering{\resizebox{8.5cm}{!}{\includegraphics{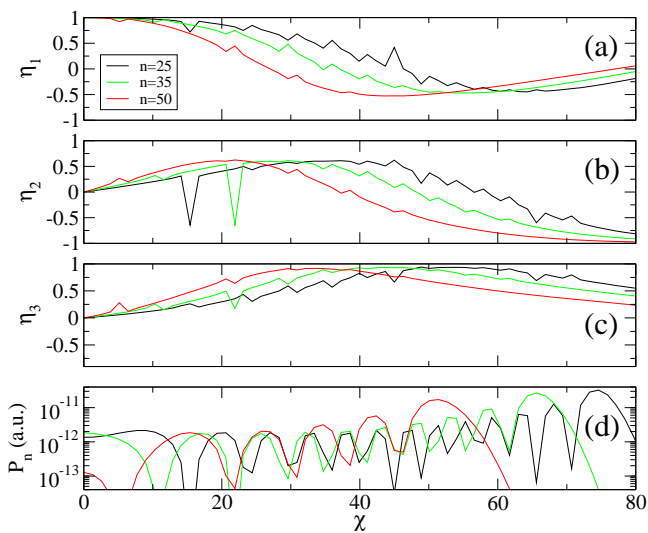}}}
\caption{\label{stokes} Panels (a)-(c). Stokes parameters $\eta_1$, $\eta_2$, $\eta_3$
as a function of the angle $\chi$ for different numbers $n$ of laser photons exchanged.
Panel (d). Differential recombination rates
$P_n=P_n[\hat{\bm \epsilon}_1(0)]+P_n[\hat{\bm \epsilon}_2(0)]$,
as a function of the angle $\chi$ for for the same values of $n$. The wavevector ${\bf k_\gamma}$
points into the direction specified by $\theta=30^\circ$,$\phi=45^\circ$.
}
\end{figure}

When ${\bf \hat{\bm \epsilon}_L}$ is parallel to ${\bf q}$, $T_n[\hat{\bm \epsilon}_2(0)]$ is zero because the dipole moment
transition matrix between states with the same magnetic quantum number has no component along
the direction perpendicular to the quantization axis, and $\eta_2$, $\eta_3$ turn out to be zero. Once ${\bf \hat{\bm \epsilon}_L}$
is no longer aligned to ${\bf q}$, a time-dependent dipole moment with components along
${\bf \hat{\bm \epsilon}_1}(0)$ and ${\bf \hat{\bm \epsilon}_2}(0)$ is induced by the driving field.
Interference between the components of the transition dipole moment along these
two perpendicular directions allows emission of photons whose circular polarization
degree depends on $\chi$ and their propagation direction.

For $\chi=\pi/2$, $\eta_3$ vanishes. This result may be easily explained by resolving the periodical part
of the dipole moment matrix element into terms of its Fourier component. From the invariance of the Hamiltonian
under the simultaneous reflection in the plane $zy$ and time translation of $\pi/\omega_L$, it may be shown
that $T_n[\hat{\bm \epsilon}_2(0)]$ is zero for processes involving exchanges of even numbers of the
laser field assisting photons, while for odd $n$, the phase difference between
$T_n[\hat{\bm \epsilon}_1(0)]$ and $T_n[\hat{\bm \epsilon}_2(0)]$ is zero. Therefore, according to
Eq.~\ref{eta3}, $\eta_3$ is found to be zero for both odd and even $n$.

Finally, we observe that by increasing $\chi$, the coupling of the electron
with the assisting radiation field weakens, leading to a decreasing of
the number of the laser photons exchanged during the recombination process.
Therefore the disalignment of the oscillating field with the incoming electron
momentum causes the narrowing of the emission spectra width.
For the channels shown in Fig.~\ref{stokes}, when $\chi$ is close to $\pi/2$ the differential emission rates
become vanishing small, therefore results for $\chi>80^\circ$ have not been shown.

By concluding, we remark that, although numerical calculations of the recombination rates have required
use of approximate wavefunctions,
the major goal of our work has been to show, on symmetry grounds, that in the presence a linear polarized laser,
during the recombination event, elliptically polarized photons are emitted, the ellipticity depending on the disalignment
between the electron beam momentum and the oscillating electric field direction. Moreover,  due to the presence of the external field,
strong modifications to the angular distribution of the emitted photons have been found.

\begin{acknowledgments} We
This work was supported in part by the Italian Ministry of University
and Scientific Research through the PRIN ``Nanostructures Photodeposition
for Nonlinear Optics'' prot.~2004023130.
\end{acknowledgments}

\end{document}